\documentclass[conference]{IEEEtran}
\IEEEoverridecommandlockouts
\usepackage{cite}
\usepackage{amsmath,amssymb,amsfonts}
\usepackage{algorithmic}
\usepackage{graphicx}
\usepackage{textcomp}
\usepackage{xcolor}
\usepackage{array}

\usepackage{booktabs}

\newcommand{\reffig}[1]{Fig.\ref{#1}}
\newcommand{\refeqs}[1]{Eq.\ref{#1}}

\def\BibTeX{{\rm B\kern-.05em{\sc i\kern-.025em b}\kern-.08em
    T\kern-.1667em\lower.7ex\hbox{E}\kern-.125emX}}
\begin{document}

\title{Dynamic Transaction Storage Strategies for a Sustainable Blockchain}

\author{\IEEEauthorblockN{1\textsuperscript{st} Zhao Xiongfei}
\IEEEauthorblockA{\textit{Department of Computer and Information Science} \\
\textit{University of Macau}\\
Macau \\
yb97480@um.edu.mo}
\and
\IEEEauthorblockN{2\textsuperscript{nd} Yain-Whar Si}
\IEEEauthorblockA{\textit{Department of Computer and Information Science} \\
\textit{University of Macau}\\
Macau \\
fstasp@um.edu.mo}
}

\maketitle

\begin{abstract}
As the core technology behind Bitcoin, Blockchain's decentralized, tamper-proof, and traceable features make it the preferred platform for organizational innovation. In current Bitcoin, block reward is halved every four years, and transaction fees are expected to become the majority of miner revenues around 2140. When transaction fee dominates mining rewards, strategic deviations such as Selfish Mining, Undercutting, and Mining Gap could threaten the integrity and security of the Blockchain. This paper proposes a set of Dynamic Transaction Storage (DTS) strategies for maintaining a sustainable Blockchain under the transaction-fee regime. We demonstrate that block incentive volatility can be reduced through systematic simulation by applying DTS strategies and avoiding strategic deviations. With DTS, public Blockchains such as Bitcoin become sustainable when the mining reward is solely based on the transaction fee.
\end{abstract}

\begin{IEEEkeywords}
Blockchain, Merkle Tree, Transaction Fees, Block Incentive, Dynamic Transaction Storage
\end{IEEEkeywords}

\section{Introduction}

The idea of Blockchain was first introduced by Satoshi Nakamoto \cite{nakamoto2008Bitcoin} in 2008 and later carried out for the first time as the public ledger for Bitcoin currency in 2009. Blockchain technology becomes one of the main drivers behind digital currencies and a potential foundation for value transfer and public ledger. It has been widely used in various fields such as financial systems, intelligent transportation, Internet of things, smart grid, voting systems, and data center networks \cite{AGGARWAL201913}, etc.

Blockchain is a distributed ledger and can be used to record all transaction information securely. The essence of this technology is that different nodes participate in a distributed database, and each node records all the transactions in the network. Transactions are packed into a block that contains a pointer to a previous block. As a result, a Blockchain is formed by linking the transaction information from different time periods in the form of connected blocks. These blocks are typically verified by nodes referred to as miners. 

The role of miners is to guarantee the security, execution, and authenticity of transactions. They group transactions waiting in the transaction pool for the network to get incorporated into a block. The miner advertises a block in the Blockchain network once it completes its mining for this new block to claim the mining reward. The newly mined block is verified by the majority of miners in the Blockchain network before it is appended to the longest chain. Mining incentive plays an essential role as a financial reward to ensure that distributed hash power is spread among enough different miners. Therefore, no one can monopolize over 51\% mining power to initiate attacks such as double-spending \cite{10.1145/2382196.2382292}.

Mining incentives include a fixed number of system-generated coins and also the user-submitted transaction fees. In Bitcoin, the coin-based reward is still the current primary revenue source, even though it is preset to be halved approximately every four years and will gradually be replaced by transaction fees. Easley et al. \cite{EASLEY201991} concluded that with increasing Bitcoin price levels, transaction fees could play only a secondary role in explaining miners' willingness to participate. Satoshi Nakamoto mentioned that "The incentive can also be funded with transaction fees. Once a predetermined number of coins have entered circulation, the incentive can transition entirely to transaction fees and be completely inflation-free" \cite{nakamoto2008Bitcoin}.

In current Bitcoin, a block creation time is, by design, around 10 minutes, and the general heuristic for accepting a transaction to be valid is when the transaction is six blocks old, which yields 60 min on average. However, as the miners are inclined to prioritize the transactions that offer higher fees, a typical transaction will take longer to be approved during congested times. To this end, although Bitcoin has opened new opportunities for value transfer and transparency against censorship, it has been long criticized for its slow transaction confirmation times and high transaction fees.

The contradiction between fast settlement and transaction fees will become more prominent in transitioning to a transaction-fee regime. Chiu et al. \cite{10.1093/rfs/hhy122} suggest that to discourage incentives to fork the chain, investors need to pledge enough transaction fees as mining rewards to maintain a sufficient amount of mining activities. Under this circumstance, Blockchain-based settlement systems need to make fast settlement a scarce resource for investors to pledge transaction fees that generate mining rewards. However, with the required transaction fee increasing, the proportion of fee-submitting users decreases slowly at first, followed by a sharp decline \cite{2020Analyzing}. Thus, the system needs to balance transaction fees and settlement speed to maximize the expected net trade surplus while ensuring that the Blockchain is tamper-proof.

The interaction of settlement speed and high transaction fee \cite{10.1007/978-3-662-48051-9_2} could also limit the scalability of Blockchain. In often argued scenarios, settlement speed is related to \textsl{block size} (how many transactions can be included in one block) and \textsl{block time} (how frequently new transactions are incorporated into Blockchain). Increasing block size improves Blockchain throughput, whereas reducing block time could improve Blockchain latency. However, one can not set arbitrarily bigger blocks, because smaller block creates congestion, which is necessary for investors to pay a higher transaction fee for early settlement. One can not set arbitrarily low block time either because lower block time will cause rewards to split over more blocks. As a result, it can lead to a reduction in the reward per block as well as the mining competition \cite{10.1093/rfs/hhy122}.

Carlsten et al. \cite{10.1145/2976749.2978408} further revealed other severe issues when Blockchain transitions to a transaction-fee regime. Specifically, the time-varying nature of transaction fees allows a richer set of strategic deviations such as \textsl{Selfish Mining} \cite{10.1007/978-3-662-45472-5_28}, \textsl{Undercutting} \cite{10.1145/2976749.2978408}, \textsl{Mining Gap} \cite{10.1145/2976749.2978408}, \textsl{Pool Hopping} \cite{DBLP:journals/corr/abs-1112-4980}, etc. These strategic deviations would not present in the block-reward model. Moreover, when misbehaving miners carry out strategies such as \textsl{Undercutting}, they seek to enhance their chance to fork the chain by giving up part of available transaction fees. When miners are not maximizing mining reward by prioritizing the transaction fee, there is also no motivation for investors to pay higher transaction fees to compete for scarce early settlement. This situation invalidates the theory described in the previous paragraph, which argues that fast settlement becomes a scarce resource.

Due to the exponentially distributed transaction arrival time with the different transaction amounts, the volatility of the block incentive is very high under the transaction-fee regime. As a result, miners may adopt different deviant strategies to deal with the situation. Carlsten et al. \cite{10.1145/2976749.2978408} argued that deviant mining strategies in a transaction-fee regime could hurt the stability of Bitcoin mining and harm the ecosystem. In order to make the Blockchain sustainable, in this paper, we utilize the fixed block size of Bitcoin to generate stable mining rewards, thus converting the time-varying transaction fees into stable block rewards under the transaction-fee regime. This fundamental thinking leads us to propose a set of novel Dynamic Transaction Storage (DTS) strategies. To this end, in this paper, we rigorously evaluate 14 different DTS strategies for stabilizing the volatility. Among all the tested strategies in the experiments, we found that the DTS strategy, which is based on a time-based transaction incorporation priority and a designated space for small transactions, is the most promising strategy in reducing the block incentive volatility. The contributions of our proposal are three folds: 

\begin{itemize}
	\item[a)] We propose a set of Dynamic Transaction Storage (DTS) strategies to minimize the volatility of mining incentives for each block. Our analysis reveals that DTS strategies can stabilize block incentive regardless of whether transactions are incorporated by arrival sequence (Time-based) or a transaction fee (Fee-based). By stabilizing mining incentives for each block under the transaction-fee regime, we could eliminate strategic deviations such as \textsl{Selfish Mining}, \textsl{Undercutting}, \textsl{Mining Gap}, \textsl{Pool Hopping}.

	\item[b)] Stabilized block incentive for Time-based transaction incorporation strategy opens up the possibility to make Blockchain scalable under the transaction-fee regime. By reducing block time, we could scale Blockchain without worrying about not having sufficient mining competition rewards. Because in current Bitcoin, it has to maintain settlement speed at a certain level to exploit investors' willingness to pledge more transaction fees. However, under the Time-based transaction incorporation mechanism, fast settlement is no longer a scarce resource. The restriction of settlement speed and high transaction fees no longer exist.

	\item[c)] In the proposed DTS approach, the size of the block that contains higher-fee transactions is smaller compared to other blocks. Therefore, blocks with higher transaction fees can reach consensus more quickly on the Blockchain. Besides, in the current Bitcoin protocol, to make fast settlement a scarce resource, block time and block size are used to limit how frequently new information is added to the Blockchain. Whereas in DTS, we take advantage of consensus latency nature to settle higher-fee transactions faster. In DTS, the scalability of Blockchain is no longer restrained by block time and block size.
\end{itemize}

This paper is organized as follows. Section 2 summarizes the related work. In Section 3, we give the Background and Preliminaries for understanding the related concepts. Section 3 describes the simulation attributes and analysis results. The paper is concluded in Section 4.

\section{Related Work}

A major success factor of Bitcoin is its novel reward mechanism. Miner's revenues include a fixed number of system-generated Bitcoins and also the user-submitted transaction fees. Since the coin-based reward is preset to be halved after every 210,000 blocks are mined and eventually can reach zero sometime around 2140 \cite{EASLEY201991}, the transaction fee will become major compensation for the miners. Because Bitcoin's issuance structure is designed to be deflationary, each block incentive depends on the Bitcoin price's market value. While block rewards in terms of the number of Bitcoins being mined are decreasing, the increasing market value of Bitcoins has ultimately resulted in a stable block reward. To this end, total revenue for each block is still expected to be sufficient to sustain the Blockchain before the coin-based reward reaches zero.

Besides the coin-based rewards, miners' transaction fee incentives also play an increasing role as the coin-based reward dwindled. For the transaction fee incentive, users typically submit transactions with a certain amount of associated fees to get their desired priority and stimulate miners to confirm their transactions preferentially. All the transactions pending confirmation are stored in the memory pool. As time progresses, there are more pending transactions in the system, and the potential fees grow. Since the block size is limited, the number of transactions that can be recorded into a block by miners is restricted. As such, revenue-maximizing miners naturally first select and pack those transactions with higher priorities as their mining basis. Bitcoin incentives are expected to transit from block-reward regime to transaction-fee regime. More and more studies are focusing on the impact of the transaction-fee regime.

Under the transaction-fee regime, Bitcoin investors hope to optimize their transaction fees to minimize their trading costs. In this context of incorporating biding mechanism, Li et al. \cite{LI2019113094} propose to adopt Generalized Second Price (GSP), which is a better biding mechanism than Generalized First Price (GFP). Under the GFP mechanism, transactions are processed in a pipeline with an order defined by the "rank-by-fee" mechanism.  Besides, the payment rule is defined by the "pay-its-bid" mechanism. Transactions with higher fees are usually confirmed first, and users need to pay their submitted fees.

However, transaction fees might reach or even exceed the trading amount, especially in micro-payment scenarios. To minimize investors' cost, GSP adopts a weighted fee to determine transactions' ranks, which can be calculated by user-submitted fees, transaction amount, quality scores, and virtual fees accumulated from the waiting time. Confirmed transactions should transfer fees to miners, which are equal to the next highest fee to miners. Under the GSP system, the transaction cost can be reduced, and the system efficiency can be improved.

From Bitcoin miners' point of view, the amount of transaction fees included in a block is linearly proportional to its size, and the payoff is also determined mainly by a miner's winning probability \cite{houy2014economics}. The expected reward increases as the block grow larger, while the winning probability goes down since smaller blocks tend to reach consensus faster. Jiang et al. \cite{8946169} suggest that in order to maximize miners' profit and to ensure that miners have no incentive to misbehave, a block size should be limited to 4 MB. Nicolas Houy \cite{houy2014economics} also suggest that one cannot set arbitrarily large block sizes because it could jeopardize the very existence of Bitcoin in the long term.

Another factor that can affect the miners' profit is the block time, representing how frequently new transactions can be incorporated into one block. In current Bitcoin, block time is around 10 minutes, which means miners will spend approximately 10 minutes to mine a new block by solving a computationally difficult problem. Chiu et al. \cite{10.1093/rfs/hhy122} suggest that one can not set arbitrarily low block time either. Decrease in block time means total rewards have to be split over more blocks, thus reducing the reward per block and the mining competition.

From the Bitcoin group welfare point of view, mining is a public good for validating individual transactions: once there is a sufficient amount of mining activities or, equivalently, transaction fees, forking can be prevented independent of the total number of transactions. Given the current Bitcoin mechanism, a miner usually has more incentives for choosing a set of transactions whose fees are very close to whatever real-valued target he/she wants. Thus, the chances are high that a transaction with a lower fee could not be processed immediately or might not even be relayed by miners \cite{10.1093/rfs/hhy122}.

Easley et al. \cite{EASLEY201991} suggest that transaction fee are not welfare-improving. Higher transaction fees can induce users to drop out, and increasing waiting times can also cause users to disengage. Meanwhile, the benefit for an investor to create a fork is related to the individual transaction amount, whereas the cost of doing so depends on the mining reward, which is related to the aggregate transaction volume \cite{10.1093/rfs/hhy122}. Once Bitcoin switches to the transaction-fee regime, the correlation between forking cost and total transaction fee in one block will become more obvious.

Carlsten et al. \cite{10.1145/2976749.2978408} further revealed other severe issues when Bitcoin mining transitions to a transaction-fee regime. Besides, the time-varying nature of transaction fees allows a richer set of strategic deviations outlined below. Therefore, it simply would not present in the block-reward model because of the high volatility of mining incentives.

\begin{itemize} 

	\item \textbf{\textsl{Selfish Mining}} Volatility mining incentives could lead to miner's misbehavior. Selfish miners hold newly mined blocks without disclosing them, which enables miners to get more than their fair share of rewards \cite{10.1007/978-3-662-45472-5_28}. Also, in a transaction-fee regime, selfish mining is immediately profitable \cite{10.1145/2976749.2978408}.

	\item \textbf{\textsl{Undercutting}} \cite{10.1145/2976749.2978408} According to Bitcoin mining protocol, each miner can decide what and how many transactions to include in their block. That also means miner can deliberately leave high-fee transactions in the mempool to attract other miners to extend their chain. If there is a 1-block fork, it is more profitable for the next miner to break the tie by extending the block that leaves the most available transaction fees rather than the oldest-seen block.

	\item \textbf{\textsl{Mining Gap}} \cite{10.1145/2976749.2978408}  Without a block reward, immediately after a block is found, there is zero expected reward for mining but nonzero electricity cost. Uncertainty in mining rewards, and even unprofitable risk, makes it difficult for miners to estimate their return on investment (ROI) of mining power and reduce their incentive to participate in mining activities.

	\item \textbf{\textsl{Pool Hopping}} \cite{DBLP:journals/corr/abs-1112-4980} Miner's expected reward for participation varies over time, depending on how many shares have been contributed since the pool found its last block. The concern is that miners would respond by "hopping" in real-time to the pool that maximizes their expected rewards. 

\end{itemize}

With the decline in numbers of Bitcoin, transaction-fee regime supposes to play an increasingly important role. Even further, after coin-based rewards reach zero, the transaction-fee regime is expected to become a major mining incentive resource to maintain the sustainability of the Blockchain. However, these studies suggest that these welfare and user participation effects are why transaction fees alone are not a panacea for the dynamic challenges facing the evolving Bitcoin Blockchain.

\section{Motivations}

For Bitcoins mining in the future, under the transaction-fee regime, there will be no steady large block reward. Instead, the transaction fee will play a major rewarding source of mining activities. In these circumstances, issues such as \textsl{Selfish Mining} \cite{10.1007/978-3-662-45472-5_28}, \textsl{Undercutting} \cite{10.1145/2976749.2978408}, \textsl{Mining Gap} \cite{10.1145/2976749.2978408}, \textsl{Pool Hopping} \cite{DBLP:journals/corr/abs-1112-4980} may affect the entire Blockchain due to the high variance in the incentives caused by the transaction fee. To make the Blockchain sustainable, we need to stabilize block incentives and avoid these deviant mining strategies. As shown in Fig.1, our main objective is to flatten the block incentive curve. In this regard, we take advantage of the fixed block size feature from the current Bitcoin design. Specifically, by dynamically allocating block space according to the transaction fees, we aim to stabilize the block incentive. To this end, in this paper, we propose a set of novel Dynamic Transaction Storage (DTS) strategies.

\begin{figure}[htbp]
	\makeatletter
	\def\@captype{figure}
	\makeatother
	\centering
	\includegraphics[width=0.4\textwidth]{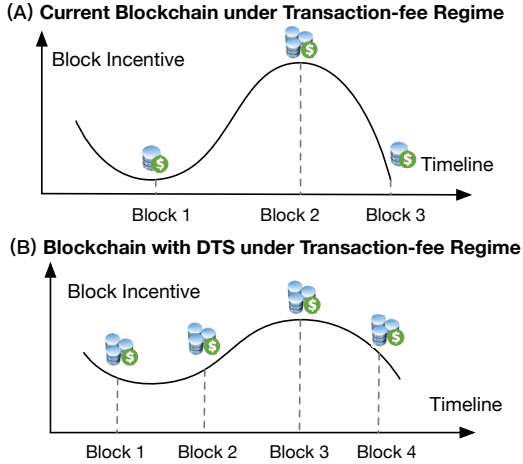}
	\caption{Variation in block incentive on a Blockchain without(A) / with(B) Dynamic Transaction Storage (DTS) Strategy}
	\label{DTS_Mechanism}
\end{figure}

\begin{figure}[htbp]
	\makeatletter
	\def\@captype{figure}
	\makeatother
	\centering
	\includegraphics[width=0.48\textwidth]{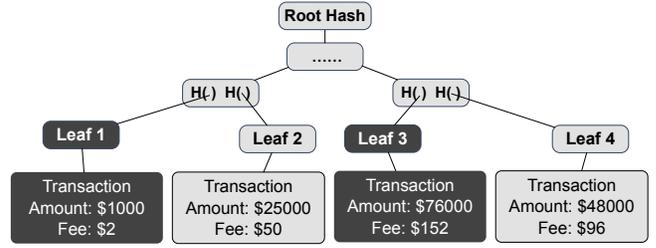}
	\caption{Constant Transaction Storage (CTS)}
	\label{MerkleTree}
\end{figure}

The design of the proposed DTS strategies is based on the storage structure of the Blockchain called Merkle tree \cite{6233691}. Suppose we have some transaction data which make up the leaves of a Merkle Tree, Merkle tree can help us efficiently verify the integrity of those transaction data. In this paper, we consider two types of storage schemes. The first storage scheme is called Constant Transaction Storage (CTS) mechanism (see \reffig{MerkleTree}). CTS is the way Bitcoin is currently used to store transactions. Transactions are stored in the Merkle tree's leaf nodes, and each leaf node corresponds to a single transaction.

In the following section, we introduce the overall idea of DTS. Next, we demonstrate that the proposed DTS strategies can stabilize block incentive volatility by simulating different transaction incorporation mechanisms and relevant attributes. Experiment results also reveal that stable block rewards can regulate miners' behavior through incentives, making Blockchain sustainable under a transaction-fee regime.

\section{Dynamic Transaction Storage (DTS) Strategies}

\begin{figure}[htbpb]
	\makeatletter
	\def\@captype{figure}
	\makeatother
	\centering
	\includegraphics[width=0.48\textwidth]{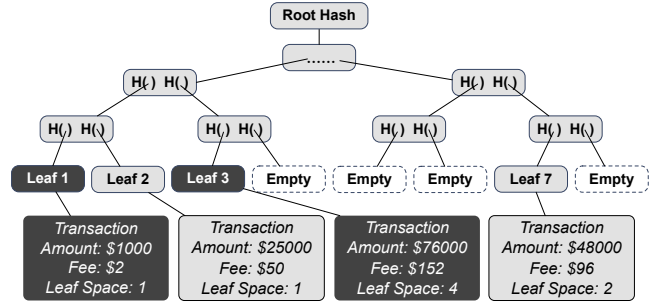}
	\caption{Dynamic Transaction Storage (DTS)}
	\label{NewMerkleTree}
\end{figure}

In this paper, we propose a set of novel Dynamic Transaction Storage (DTS) strategies based on the CTS currently used in the Bitcoin protocol. \reffig{NewMerkleTree} illustrates how DTS works. First, based on the transaction fee offered by each transaction, we calculate the space it would occupy using a Cumulative Distribution Function (CDF). Then, based on the number of leaves each transaction can occupy, we store the transactions in one of the leaves and leave the rest empty. This process is repeated for all transactions until the predefined threshold of block space for the transactions is reached.

The CDF describes the probability that a random variable $X$ with a given probability distribution will be found at a value less than or equals to $x$. We use CDF of the log-normal distribution $F(x)$ with $\sigma=6.8$ and $\mu=1$ to calculate corresponding transaction-fee $x$ based space factor (see \refeqs{con:cdf}). 

{\setlength\abovedisplayskip{0pt}
\setlength\belowdisplayskip{1pt}
\begin{equation}
F(x) = Pr(X\leq x) = \frac{1}{2} \, + \, \frac{1}{2} \text{erf} \bigg[\frac{ln(x)-\mu}{\sigma\sqrt{2}} \bigg] \label{con:cdf}
\end{equation}}

To limit the upper boundary of block incentives, we dynamically control the number of transactions incorporated based on transaction fees. We model the maximum space that each transaction could occupy as a variable denoted $m$ with cumulative distribution function (CDF). Since CDF has a maximum value of 1, the upper boundary $P_{UB}$ of transaction-fee based space factor is approximately equal to $m$ (see \refeqs{con:upperboundary}). In this way, the storage space for each transaction can be reasonably allocated, even if the transaction fee is very high.

{\setlength\abovedisplayskip{0pt}
\setlength\belowdisplayskip{1pt}
\begin{equation}
P_{UB} = \lim_{x\to\infty} F(x) \times m = m \label{con:upperboundary}
\end{equation}}

Assume $A$ denotes the transaction amount, and there are $n$ transactions included in the block. We can define $I$ to be the total BTC worth of incentive of this block. Since we set transaction fee percentage at 0.2\%, we can calculate the block total incentive $I$ using \refeqs{con:totalincentive}. The number of transactions $n$ which can be incorporated in one block is limited to 2,100 storage units.

{\setlength\abovedisplayskip{0pt}
\setlength\belowdisplayskip{1pt}
\begin{equation}
I=\sum_{i=1}^n A_n \times 0.2\%  \label{con:totalincentive} 
\end{equation}}

Note that the number of transactions $n$ which can be incorporated in one block is limited to 2,100 storage unit. This constraint can be expressed as follows:

{\setlength\abovedisplayskip{0pt}
	\setlength\belowdisplayskip{1pt}
	\begin{equation}
		\sum_{i=1}^n F(A_n \times 0.2\%) \times m < 2100
\end{equation}}

\subsection{Attributes Considered in DTS Strategies}

In the proposed set of DTS strategies, we consider four main factors (attributes). Although we could consider additional factors in designing the strategies, we aim to minimize the number of factors in achieving the objectives stated in Section 1. Simplifying the number of factors considered for the DTS strategies also allows us to better understand the impact of each attribute on the outcome of the experiments. The factors considered in the proposed DTS strategies are as follows: 

\begin{itemize} 
	
	\item \textbf{Transaction Storage Mechanism (A1):} In the proposed DTS strategies, we consider two types of transaction storage schemes namely CTS and DTS.  \textsl{Constant Transaction Storage (CTS)} means one transaction can take up one leaf space of Merkle Tree as current Bitcoin. Whereas \textsl{Dynamic Transaction Storage (DTS)} means that instead of storing the transaction in one leaf, we store it using multiple leaves. How many leaves one transaction can occupy depends on how much transaction fee this transaction provides.
	
	\item \textbf{Transaction Incorporation Priority (A2):} In DTS strategies, we consider Time-based and Fee-based transaction incorporation mechanisms instead of allowing miners to choose which and how many transactions to incorporate into blocks. For \textsl{Time-based}, transactions are incorporated into blocks on a first-come, first-served basis. For \textsl{Fee-based}, transactions with higher fees will be incorporated into blocks first.
	
	If there are $R$ transaction fees available in current Bitcoin, the miner can choose to include any real-valued number of transaction fees between 0 and $R$ in their block. That is, transactions are fine-grained enough that a miner can selectively choose a set of transactions whose fees are very close to whatever real-valued target they have in mind. Although, miner's ability to select transactions is the main factor causing Undercutting attacks. 
		
	\item \textbf{Designated Space for Small Transactions (A3):} In the early stages of Bitcoin, 50KB in each block was reserved for high-priority transactions without fees. That means the transaction confirmations are conducted in two separate pipelines, where the one is for transactions with fees, and the other is for priority transactions without fees \cite{LI2019113094}. When the arrival rate of potential transactions increases, this mechanism can avoid only transactions with fees attached are posted to the Blockchain.
	
	In our experiments, we extend this parameter's scope, and we designate block space for all the transactions which fees are below a certain threshold. By introducing this attribute, we can avoid only transactions with higher fees are incorporated into blocks under the Fee-based transaction incorporation mechanism. The benefit of this parameter goes beyond solving the Fee-based transaction selection dilemma. It also helps both the Fee-based and Time-based transaction incorporation mechanism to control small transaction incorporation percentages per block. To this end, we designate 100 units of block space for a transaction fee lower than 1.5 dollars. Therefore, the minimum reward level of each block is limited by: \begin{equation}(2100-100)\times1.5=3000\end{equation}
	
	\item \textbf{Maximum Space for One Transaction (A4):} For CTS mechanism, one transaction can only occupy one leaf space of Merkle tree. For the proposed DTS strategies, a transaction with a higher fee will be incorporated into a block by occupying multiple leaf spaces. We adopt the Cumulative Distribution Function (CDF) model to allocate leaf space based on the transaction fee dynamically. This attribute controls the maximum storage space for each transaction and allows transactions to be reasonably allocated, even with an arbitrarily large transaction fee.
	
\end{itemize} 

\subsection{Benchmark}

In finance, volatility represents the variation to which price series change over time \cite{GLOSTEN198571}. We use volatility as the simulation benchmark to measure the fluctuation degree of block incentive. Specifically, volatility is a measure of the uncertainty of mining activities' return rate and can reflect the risk level of mining activity. High volatility means that the block incentive varies dramatically from block to block. Under these circumstances, the uncertainty of the mining reward is also very high. Low volatility means block incentive changes minimally, miners tend to benefit from low volatility because they can secure stable returns from mining.

The following strategic deviations are directly related to the volatility of block incentives. Therefore, in this paper, we use volatility to evaluate the proposed DTS strategies.

\begin{itemize}
	
	\item \textit{Selfish Mining} strategy means miners holding newly mined blocks without disclosing them until more high-fee transactions arrive at the mempool and are included in the block. Carlsten et al. \cite{10.1145/2976749.2978408} propose an Improved Selfish-Mine strategy. In that strategy, under the transaction-fee regime, selfish miners can base on block incentives to decide whether to hide their blocks. This enables selfish mining to outperform both default mining and traditional selfish mining. If the volatility of block incentive is high, it will give selfish miner the opportunity to determine the optimal cutoff time.
	
	\item In \textit{Undercutting}, deviant miners could intentionally give up part of the available transaction fees. Those unclaimed transactions (associated with fees) can incentivize more strategic miners to fork the chain by extending attacker's block to gain more rewards. But this also means the block reward to extend older block or extend fork has certain difference for the strategic miners. Therefore, the block incentive volatility may enable deviant miners and strategic miners to initiate undercutting strategy.
	
	\item During Bitcoins mining, the miners bear the costs of electric power to run their mining rigs.  In \textit{Mining Gap}, without a block reward, immediately after a block is found, there is zero expected reward for mining but nonzero electricity cost, making it unprofitable for any miner to mine. With high block incentive volatility, the uncertainty over whether mining activities can be profitable is very high.
	
	\item In \textit{Pool Hopping}, the block incentive is directly related to the transaction amount in the mempool. Miners tend to mine on the mempool, which has more and bigger transactions to earn more block rewards. Miners will compete to incorporate bigger transactions in the mempool to maximizes their rewards. Therefore, high block incentive volatility can make miner's expected reward for participation varies over time.
	
\end{itemize}

The standard deviation of logarithmic returns can be used to measure the volatility \cite{doi:10.2469/faj.v46.n3.23}. In this paper, we use a similar approach to measure the fluctuation of block incentive. Assume that $n$'s block total incentive is $I_n$. By using  $I_n$, we can calculate $R_{n}$	as follows:
\setlength\abovedisplayskip{2pt}
	\setlength\belowdisplayskip{2pt}
	\begin{equation}
		R_{n}=ln(I_{n}/I_{n-1})\label{reward}
\end{equation}

To calculate the standard deviation of block incentive, we need to calculate the average of $R_{n}$. The average of continuously compounded block incentive $R_{avg}$ can be defined as follows:
\setlength\abovedisplayskip{2pt}
	\setlength\belowdisplayskip{2pt}
	\begin{equation}
		R_{avg}=\frac{\sum_{i=1}^nR_{i}}{n}\label{average} \vspace{5ex}
\end{equation}

$\sigma$ is the standard deviation of block incentive in a particular period. By using a rolling window - a period of $n$ consecutive blocks ending on the last block, we can calculate $\sigma$ as follows:
\setlength\abovedisplayskip{2pt}
	\setlength\belowdisplayskip{0pt}
	\begin{equation}
		\sigma = \sqrt{\frac{\sum_{i=1}^n(R_{i}-R_{avg})^{2}}{n-1}}\label{standard deviation}
\end{equation}

\section{Experiment Settings}

In our experiments, we use SimBlock \cite{8751431} as our simulation platform. It allows us to simulate a large-scale peer-to-peer network of a public Blockchain such as Bitcoin \cite{nakamoto2008Bitcoin} and Litecoin \cite{litecoin}. SimBlock can be configured with parameters such as hash power distribution, network latencies, and incentive mechanisms, etc. In the default setting, blocks are generated by a probability assuming Proof-of-Work and propagated along with the simulated Blockchain network. It can also be applied to the public Blockchain that retains other mining mechanism schemes such as Proof-of-Stake (PoS) \cite{vasin2014blackcoin}, Proof-of-Burn (PoB) \cite{10.1007/978-3-030-51280-4_28}, etc.

Based on the current running state of Bitcoin, we list the assumptions for our experiment in Table~\ref{tab1}. We use Bitcoin historical market data for the period of Dec 2019 to September 2020. These historical data contains 400,000 transactions with minute-to-minute updates of OHLC (Open, High, Low, Close), volume in BTC, and weighted Bitcoin price \cite{BitcoinHistoricalData}. The Bitcoin protocol itself has requirements on the block size. A block should be bounded within 1 MB and includes all valid transactions if available transactions are not enough to occupy a 1 MB block fully. Different transactions contain different sizes of information, about 500 bytes on average. We take Bitcoin's one-year average value of transactions per block \cite{TransactionsPerBlock}, that is, about 2,100 transactions per block.

\begin{figure}[htbp]
	\makeatletter
	\def\@captype{figure}
	\makeatother
	\centering
	\includegraphics[width=0.44\textwidth]{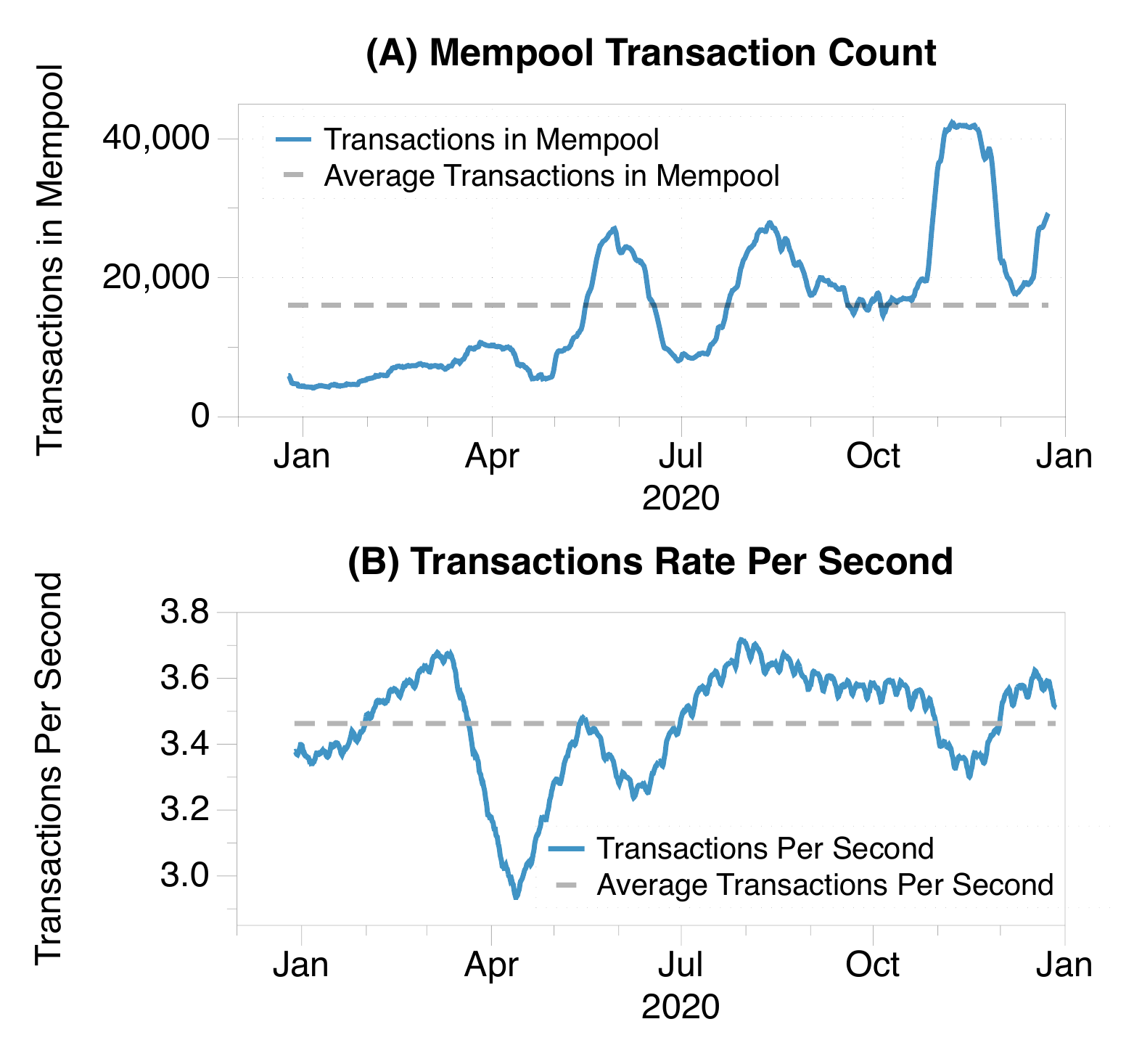}
	\caption{Mempool Transaction Count(A) and Transaction Rate Per Second(B)}
	\label{MempoolTrxCount_TrxRatePerSec}
\end{figure}

To facilitate our experiment, we also alter SimBlock to add mempool model with a constant arrival rate of transactions to SimBlock. Mempool \cite{Mempool} is the name given to the set of valid transactions that the miner is aware of but has not yet been included in a block. As illustrated in \reffig{MempoolTrxCount_TrxRatePerSec}, we take the one-year average value of Bitcoin mempool transaction volume, that is, 16,000 transactions, as the initial value of mempool. Based on the one-year average value of transaction rate per second \cite{TransactionsRatePerSecond}, we set 3.5 new transactions to arrive at mempool every second. Bitcoin consensus algorithm sets the difficulty level such that, on average, a block is added to the Blockchain every ten minutes.

 Under the current Bitcoin protocol, revenue-maximizing miners tend to prioritize those transactions with higher fees, forcing Bitcoin users to increase their transaction fees for faster confirmation or otherwise queue and wait. Regardless of whether GFP or GSP mechanism is used as a transaction fee biding mechanism, they may reduce users' costs from a single transaction point of view. From a block incentive point of view of the transaction-fee regime, no matter which transaction fee biding mechanism is used, it can not avoid deviant miner behavior. To this end, in our experiments, transaction fees are set for a fixed percentage (0.2\%) of the transaction amount to simplify our model.

\linespread{1.3}
\begin{table}[htbp]
\caption{Simulation settings based on the Blockchain Protocol}
\begin{center}
\begin{tabular}{|c|c|}
\hline
\textbf{Parameter} & \textbf{\textit{Value}} \\
\hline
Transaction Size & 500 Byte \\
\hline
Block Size & 1 MB \\
\hline
Transactions for One Block & 2,100 Transactions \\
\hline
Transactions in Mempool & 16,000 Transactions \\
\hline
New Transaction Rate & 3.5 Transactions per Second \\
\hline
Transaction Fee Precentage & 0.2\% \\
\hline
\end{tabular}
\label{tab1}
\end{center}
\end{table}

\section{Experiment Results}

We consider two sets of experiments, Experiment 1 includes eight strategies (Strategy 1, Strategy 2,..., Strategy 8) based on the combination of attributes A1, A2, A3, and A4. Experiment 2 includes six strategies (Strategy 9, Strategy 10,..., Strategy 14) based on attributes A2 and A4. Experiment 1 is mainly used to evaluate the effectiveness of DTS in reducing block incentive volatility. After confirming that DTS can effectively reduce the block incentive volatility, we further verify the effectiveness of DTS by using Experiment 2's strategies with different maximum spaces for one transaction.

\subsection{Experiment 1: Strategies With/Without DTS}

\begin{figure}[htbp]
	\makeatletter
	\def\@captype{figure}
	\makeatother
	\centering
	\includegraphics[width=0.38 \textwidth]{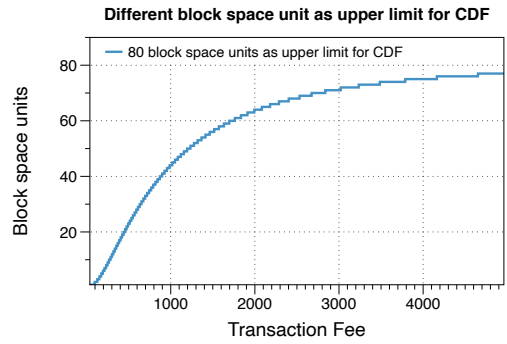}
	\caption{CDF with maximum 80 leaves for One Transaction}
	\label{CDF80}
\end{figure}

In Experiment 1, the first 4 strategies Strategy 1, Strategy 2, Strategy 3, Strategy 4 are for the Constant Transaction Storage (CTS) mechanism and the maximum space for one transaction is set to 1 block. The second group of strategies, Strategy 5, Strategy 6, Strategy 7, and Strategy 8 are for Dynamic Transaction Storage (DTS) mechanism, and the maximum space for one transaction is set to 80 blocks.  As shown in \reffig{CDF80}, the relationship between transaction fees and block space unit is calculated according to the CDF algorithm. A transaction needs to occupy at least one block space unit, and can occupy up to 80 block space units based on its transaction fee. Therefore one transaction is set to 80 blocks in the experiments. All the strategies considered in Experiment 1 are listed in Table ~\ref{tab2}.

\begin{table}[htbp]
\caption{Strategies in Experiment 1}
\begin{center}
\begin{tabular}{|c|c|c|c|c|}
\hline
\textbf{Strategy} & \textbf{\textit{A1}} & \textbf{\textit{A2}} & \textbf{\textit{A3}}  & \textbf{\textit{A4}} \\
\hline
Strategy 1 & Constant &  Time-based & No & 1 \\
\hline
Strategy 2 & Constant &  Time-based & Yes & 1 \\
\hline
Strategy 3 & Constant &  Fee-based & No & 1 \\
\hline
Strategy 4 & Constant &  Fee-based & Yes & 1 \\
\hline
Strategy 5 & Dynamic & Time-based & No & 80 \\
\hline
Strategy 6 & Dynamic & Time-based & Yes & 80 \\
\hline
Strategy 7 & Dynamic & Fee-based & No & 80 \\
\hline
Strategy 8 & Dynamic & Fee-based & Yes & 80 \\
\hline
\multicolumn{4}{l}{A1 - Transaction Storage Mechanism}\\
\multicolumn{4}{l}{A2 - Transaction Incorporation Priority}\\
\multicolumn{4}{l}{A3 - Designated Space for Small Transaction}\\
\multicolumn{4}{l}{A4 - Maximum Space for One Transaction}
\end{tabular}
\label{tab2}
\end{center}
\end{table}

\begin{figure}[htbp]
		\makeatletter
		\def\@captype{figure}
		\makeatother
		\centering
		\includegraphics[width=0.48 \textwidth]{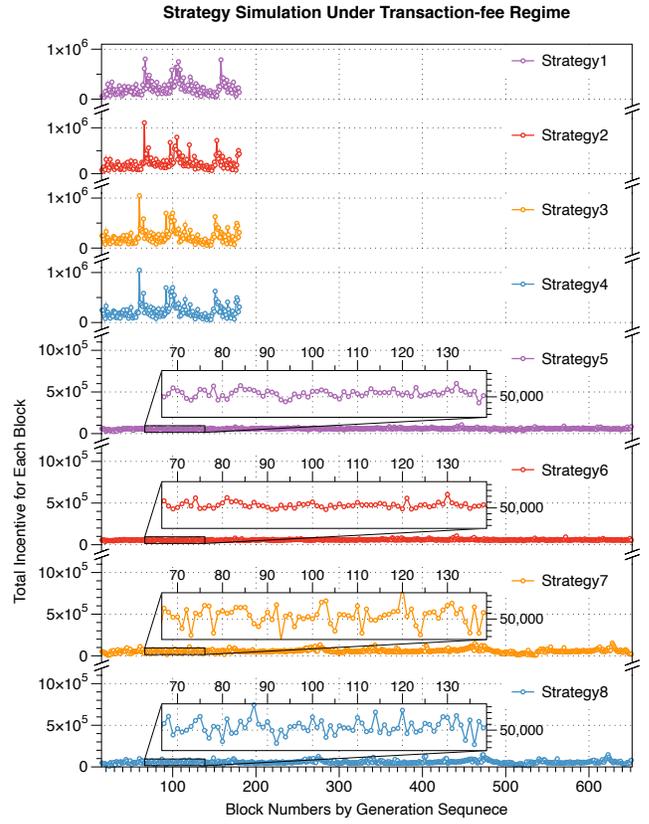}
		\caption{Total incentive of each block for the strategies in Experiment 1}
		\label{StrategySet1}
	\end{figure}

\begin{figure}[htbp]
		\makeatletter
		\def\@captype{figure}
		\makeatother
		\centering
		\includegraphics[width=0.5 \textwidth]{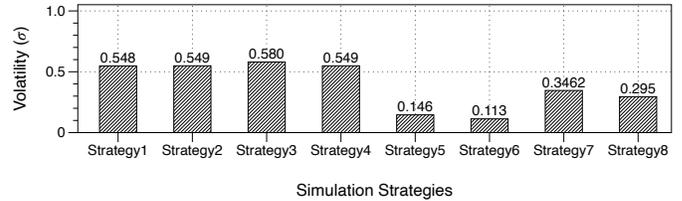}
		\caption{Comparison of volatilities for the strategies in Experiment 1}
		\label{StrategySet1Volatility}
	\end{figure}

	The results of the simulation experiments for the strategies in Experiment 1 are shown in \reffig{StrategySet1} and \reffig{StrategySet1Volatility}. In general, as the number of transactions stored in a block decreases, the Dynamic Transaction Storage (DTS) strategy slows down the incorporation process and consumes more blocks. It also causes transactions queuing in the mempool for a longer time. Due to this reason, Strategy 5-8's are longer than others in \reffig{StrategySet1}.

	In Strategy 1 and Strategy 2, the transaction incorporation mechanism is "Time-based", transactions are incorporated based on their arrival sequence. Compared to Strategy 1, because storage space for small transactions is limited, Strategy 2 has the opportunity to concentrate more large transactions into the same block. This will further increase the volatility of the block incentive of Strategy 2. As \reffig{StrategySet1Volatility} shows, the volatility of Strategy 1 and Strategy 2 is high. In addition, Strategy 2's volatility is also higher than Strategy 1.

	Strategy 3 and Strategy 4 are similar to the current Bitcoin incentive mechanism. Under the "Fee-based" transaction incorporation mechanism, transactions with higher fees will be incorporated first. In current Bitcoin, investors can either accept longer lags to save trading costs or pay a higher fee to ensure additional benefits. Our fixed percentage fee-charging assumption aligns with investors' willingness, meaning that higher fees is directly related to the transaction amount.  However, for allocating a designated space for small transactions, its impact to total block incentive is insignificant. As \reffig{StrategySet1Volatility} shows, in Strategy 4, small transactions take up a small portion of the block space that would otherwise be occupied by large transactions, and overall block incentive is slightly reduced. In Strategy 3 and Strategy 4, each block's total incentive fluctuates wildly, and Strategy 4's volatility is slightly lower.

	For Strategy 5 and Strategy 6, transactions are incorporated by their arrival sequence. Even transactions are incorporated by the arrival sequence, and investors are still motivated to pledge higher transaction fees. Because under a Time-based mechanism, even if a transaction is not prioritized at the incorporation phase due to the higher fees, the smaller size of the block containing the higher fee transactions makes the transaction faster to be settled on the Blockchain. 

	In addition to DTS being able to lower the upper limit for block incentives, we can also raise the lower limit for block rewards by introducing designated space for small transactions. From \reffig{StrategySet1Volatility} we can see that Strategy 5 and Strategy 6's volatility are much lower because DTS has successfully lower the upper boundary of block incentive. Furthermore, in Strategy 6, small transaction incorporation is conducted in a separate pipeline, and the number of small transactions in each block has been limited. This mechanism can postpone incorporating small transactions, and thus lower boundary of block incentive level has also been lifted. As a result, we can observe that Strategy 6's volatility is lower than Strategy 5.

	Strategy 7 and Strategy 8 adopt the "Fee-based" transaction incorporation mechanism. From \reffig{StrategySet1Volatility}, we can observe that DTS can reduce the block incentive volatility for both strategies.

	In summary, \reffig{StrategySet1Volatility} reveals that Strategy 1--4 has high volatility, which implies the high uncertainty in mining activities' return rate. By comparison, we can successfully reduce block incentive volatility with DTS strategies. Through our simulation, the best result is Strategy 6, in which volatility has been reduced to 0.113. Although the volatility of Strategy 8 is not as low as Strategy 6, under the current Bitcoin "pay-its-bid" mechanism, Strategy 8 can be considered as a promising solution for designing a sustainable Blockchain.

\subsection{Experiment 2: Strategies With Different Maximum Space for One Transaction}

In Experiment 2, we consider 6 strategies (Strategy 9, Strategy 10, .. , Strategy 14). In these strategies, the Transaction storage mechanism (A1) is set to "Dynamic" and the designated space for a small transaction (A3) is set to "Yes". The main purpose of these settings is to investigate how different maximum spaces for one transaction can affect the block reward volatility. These strategies are listed in Table ~\ref{tab3}.

\begin{figure}[htbp]
	\makeatletter
	\def\@captype{figure}
	\makeatother
	\centering
	\includegraphics[width=0.38 \textwidth]{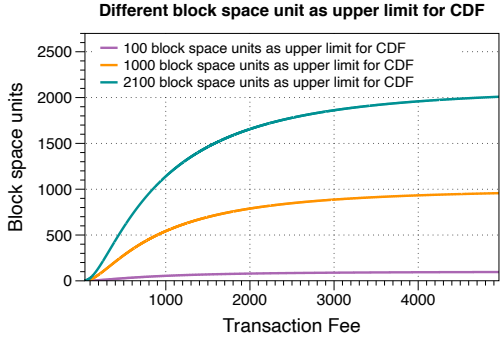}
	\caption{Comparison of (a) minimum, (b) median and (c) maximum block space unit as upper limit for CDF.}
	\label{CDF8010002100}
\end{figure}

In \reffig{CDF8010002100}, we compare the block space unit that different transactions can occupy according to their transaction fee. CDF algorithm is used to calculate the correspondence between transaction fees and different block space units. We set 100, 1000, and 2,100 respectively to simulate the minimum, median and maximum value of block space as the upper limit of one transaction can occupy.

\begin{table}[htbp]
\caption{Strategies in Experiment 2}
\begin{center}
\begin{tabular}{|c|c|c|c|c|}
\hline
\textbf{Strategy} & \textbf{\textit{A1}} & \textbf{\textit{A2}} & \textbf{\textit{A3}}  & \textbf{\textit{A4}} \\
\hline
Strategy 9 & Dynamic & Time-based & Yes & 100 \\
\hline
Strategy 10 & Dynamic & Time-based & Yes & 1000 \\
\hline
Strategy 11 & Dynamic & Time-based & Yes & 2100 \\
\hline
Strategy 12 & Dynamic & Fee-based & Yes & 100 \\
\hline
Strategy 13 & Dynamic & Fee-based & Yes & 1000 \\
\hline
Strategy 14 & Dynamic & Fee-based & Yes & 2100 \\
\hline
\multicolumn{4}{l}{A1 - Transaction Storage Mechanism}\\
\multicolumn{4}{l}{A2 - Transaction Incorporation Priority}\\
\multicolumn{4}{l}{A3 - Designated Space for Small Transaction}\\
\multicolumn{4}{l}{A4 - Maximum Space for One Transaction}
\end{tabular}
\label{tab3}
\end{center}
\end{table}

\begin{figure}[htbp]
		\makeatletter
		\def\@captype{figure}
		\makeatother
		\centering
		\includegraphics[width=0.48 \textwidth]{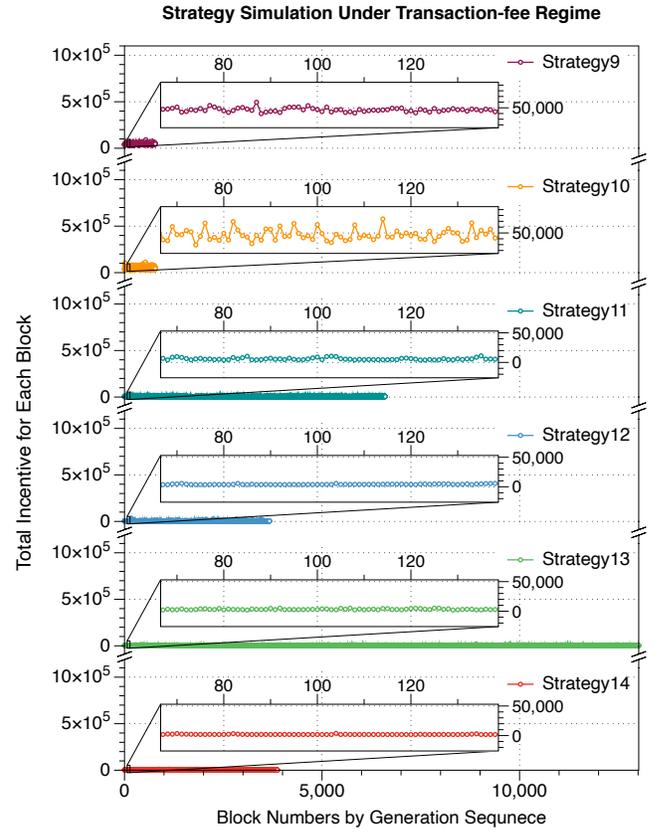}
		\caption{Total incentive of each block for strategies in Experiment 2}
		\label{StrategySet3}
	\end{figure}

\begin{figure}[htbp]
		\makeatletter
		\def\@captype{figure}
		\makeatother
		\centering
		\includegraphics[width=0.5 \textwidth]{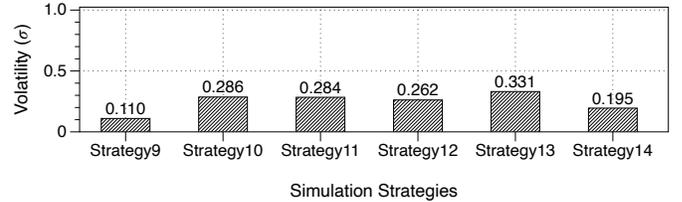}
		\caption{Volatility of Strategies in Experiment 2}
		\label{StrategySet3Volatility}
	\end{figure}

The results of the simulation experiments are shown in  \reffig{StrategySet3} and \reffig{StrategySet3Volatility}. From \reffig{StrategySet3}, we can get even lower volatility by maximizing the number of block space units for one transaction. However, after the number of block space units for a single transaction is extended to a certain point, volatility will increase. We can also observe that with the same amount of transactions, as the number of storage units a transaction can occupy increases, block consumption also increases, and the block incentive level decreases.

According to \reffig{StrategySet3Volatility}, we can observe that the magnitude of block incentive volatility in each block varies according to the CDF algorithm's parameters. The more block space unit to hold one transaction, the volatility is higher. Therefore, it is necessary to combine other variables to reduce volatility. By increasing the storage unit of one transaction alone may not necessarily the right approach.

\subsection{Analysis of Simulation Results}

For all the 14 strategies evaluated, we found that Strategy 9 is the most promising approach to reduce block incentive volatility. In the following paragraphs, we summarize the impact of DTS strategies on security and scalability properties.

\subsubsection{Security}

The security of Bitcoin's consensus protocol relies on miners behaving correctly. They are incentivized to do so via mining rewards. Under the transaction fee regime, the cost for revoking a trade depends on the mining reward, which is related to the aggregate transaction volume in one block. However, under DTS strategies, the cost for revoking a trade is related to the low volatility block incentive. As long as the incentive level is attractive enough to create enough competition for miners, the forking is less likely to happen. The stable block incentive does not rely on a predetermined number of cryptocurrencies, and it is entirely inflation-free. It is a more sustainable solution for Blockchain under the transaction-fee regime.

Any deviant miner behavior that outperforms the default is thus a serious threat to the security of Bitcoin. By lowing the volatility of block incentive, we summarize the possible prevention of the existing strategic deviation as follow:

\begin{itemize}

	\item For Improved Selfish-Mine, through stabilized block incentive volatility, the difference in block incentive is small. Therefore, selfish miners can not decide whether to hide their first block or not base on the value of the block, thus preventing the problem of Improved Selfish-Mine.

	\item For Undercutting, we simulate both Time-based and Fee-based transaction incorporated mechanism. For Time-based mechanism, miner is restricted to choose transactions in mempool based on their arrival sequence by the Blockchain protocol. Thus deviant miner can not deliberately leave unclaimed rewards for the next strategic miner. For the Fee-based mechanism, even a deviant miner leaves the unclaimed reward, and the following strategic miner can not maximize his rewards by extending the deviant miner's chain. The reason behind is that, with low block incentive volatility, the difference in block incentive is insignificant regardless of whether the longest chain or deviant miner's chain is extended. To this end, undercutting strategy can be prevented for both mechanisms.

	\item For Mining Gap, with low block incentive volatility, miners can claim the expected reward for each block in order to ensure that their constant mining power investment is profitable. This could eliminate the Mining Gap problem fundamentally and also ensure effective hash power in the Blockchain network. It can also prevent a malicious miner to fork.

	\item For Pool Hopping, by reducing the block incentive volatility, miner's reward for mining activates is still stable even when the mempool has more and bigger transactions. Mining activity's return rate is independent of the available transaction fees in the mempool, stopping miners from jumping between different mempools to maximize their reward.

\end{itemize}

\subsubsection{Scalability}

By adopting the DTS strategies, we could scale Blockchain under the transaction-fee regime by increasing the block size or setting a shorter block time. The current Bitcoin protocol is designed to create congestion to generate fees by limiting block time and block size. Bitcoin protocol designers can not arbitrarily increase the block size because doing so can slow down the prorogation speed and affect the investor's willingness to pay for the earlier settlement. Designers cannot set arbitrarily low block time either because total rewards have to be split over more blocks, reducing the reward per block.

Whereas, under the DTS strategies, the block size is no longer fixed. For a block that contains higher-fee transactions, its size is relatively smaller. When we increase the maximum block size to store more transactions, the prorogation speed will not be significantly affected for these blocks. Blocks that contain higher-fee transactions can reach consensus more quickly on the Blockchain. Even if the owners of these high-value transactions have high motivation to fork, their forking cost is relatively higher. For blocks containing lower-fee transactions, their large size causes low prorogation speed. But this does not compromise the integrity of the Blockchain and the owners of those transactions are also less likely to fork because there is little to gain. This would make the Blockchain scalable and sustainable under the transaction-fee regime.

\section{Conclusion}

This paper reviews current Bitcoin's reward mechanism and how transactions are incorporated into the block through consensus mechanisms. We list the sustainability dilemma faced by the Bitcoin designers when the block reward mechanism is switched to transaction fee-based mechanism. In recent years, Bitcoin is being criticized for slow transaction confirmation times and high transaction costs and the deviant strategies that can occur when switching to the transaction-fee regime, such as Selfish Mining, Undercutting, Mining Gap, and Pool Hopping. We argue that the root cause of these fatal strategic deviations is the high volatility of block rewards. To this end, a set of novel DTS strategies is proposed to alleviate this problem.

We show that through the systemic simulation, we could mitigate the block incentive volatility under either Time-based or Fee-based transaction incorporation mechanism by dynamically assigning Merkle Tree storage units based on the transaction fee. Among the 14 strategies evaluated, we found that strategy 9, based on a time-based transaction incorporation priority and a designated space for small transactions, is the most promising strategy in reducing the block incentive volatility. With low volatility and stable incentive fluctuation, transaction fees can play a major resource of block incentive instead of merely acting as an incentive for miners to include transactions in their blocks.

Based on the proposed dynamic allocation mechanism of block space, we discuss the difference in speed at which transactions of different amounts settle on the Blockchain. The advantages of DTS strategies to improve the scalability of Blockchain are compared with traditional Blockchain. We conclude that by adopting the DTS strategies, we can scale Blockchain by increasing the block size or by reducing the block time without affecting the security and integrity of the Blockchain. On the contrary, traditional Blockchain's scalability is restricted, and setting a large block size or shortening the block time could damage the Blockchain's security and integrity.

Although we have found a feasible way to reduce the block incentive volatility in our approach, several variables can be further investigated. We hope that this study could provide a new perspective to find a way to prevent deviant miner behavior and eliminate the severe threat to the security of Bitcoin under the transaction-fee regime. The profit distribution dilemma of private Blockchains is similar to that of Bitcoin under the transaction-fee regime. By adopting our proposed mechanism, businesses running on private Blockchain can innovatively combine their revenue model with Blockchain's mining activities' profit distribution.

\bibliographystyle{IEEEtran}
\bibliography{IEEEabrv,IEEEexample}

\begin{thebibliography}{10}
\providecommand{\url}[1]{#1}
\csname url@samestyle\endcsname
\providecommand{\newblock}{\relax}
\providecommand{\bibinfo}[2]{#2}
\providecommand{\BIBentrySTDinterwordspacing}{\spaceskip=0pt\relax}
\providecommand{\BIBentryALTinterwordstretchfactor}{4}
\providecommand{\BIBentryALTinterwordspacing}{\spaceskip=\fontdimen2\font plus
\BIBentryALTinterwordstretchfactor\fontdimen3\font minus
  \fontdimen4\font\relax}
\providecommand{\BIBforeignlanguage}[2]{{%
\expandafter\ifx\csname l@#1\endcsname\relax
\typeout{** WARNING: IEEEtran.bst: No hyphenation pattern has been}%
\typeout{** loaded for the language `#1'. Using the pattern for}%
\typeout{** the default language instead.}%
\else
\language=\csname l@#1\endcsname
\fi
#2}}
\providecommand{\BIBdecl}{\relax}
\BIBdecl

\bibitem{nakamoto2008Bitcoin}
S.~Nakamoto, ``{Bitcoin: A Peer-to-Peer Electronic Cash System},'' p.~9, 2008.

\bibitem{AGGARWAL201913}
\BIBentryALTinterwordspacing
S.~Aggarwal, R.~Chaudhary, G.~S. Aujla, N.~Kumar, K.-K.~R. Choo, and A.~Y.
  Zomaya, ``{Blockchain for Smart Communities: Applications, Challenges and
  Opportunities},'' \emph{Journal of Network and Computer Applications}, vol.
  144, pp. 13--48, 2019. [Online]. Available:
  \url{https://www.sciencedirect.com/science/article/pii/S1084804519302231}
\BIBentrySTDinterwordspacing

\bibitem{10.1145/2382196.2382292}
\BIBentryALTinterwordspacing
G.~O. Karame, E.~Androulaki, and S.~Capkun, ``{Double-Spending Fast Payments in
  Bitcoin},'' in \emph{Proceedings of the 2012 ACM Conference on Computer and
  Communications Security}, ser. CCS '12.\hskip 1em plus 0.5em minus
  0.4em\relax New York, NY, USA: Association for Computing Machinery, 2012, p.
  906–917. [Online]. Available: \url{https://doi.org/10.1145/2382196.2382292}
\BIBentrySTDinterwordspacing

\bibitem{EASLEY201991}
\BIBentryALTinterwordspacing
D.~Easley, M.~O'Hara, and S.~Basu, ``{From Mining to Markets: The Evolution of
  Bitcoin Transaction Fees},'' \emph{Journal of Financial Economics}, vol. 134,
  no.~1, pp. 91--109, 2019. [Online]. Available:
  \url{https://www.sciencedirect.com/science/article/pii/S0304405X19300583}
\BIBentrySTDinterwordspacing

\bibitem{10.1093/rfs/hhy122}
\BIBentryALTinterwordspacing
J.~Chiu and T.~V. Koeppl, ``{{Blockchain-Based Settlement for Asset
  Trading}},'' \emph{The Review of Financial Studies}, vol.~32, no.~5, pp.
  1716--1753, 04 2019. [Online]. Available:
  \url{https://doi.org/10.1093/rfs/hhy122}
\BIBentrySTDinterwordspacing

\bibitem{2020Analyzing}
\BIBentryALTinterwordspacing
J.~Li, Y.~Yuan, and F.-Y. Wang, ``{Analyzing Bitcoin Transaction Fees Using A
  Queueing Game Model},'' \emph{Electronic Commerce Research}, 2020. [Online].
  Available: \url{https://doi.org/10.1007/s10660-020-09414-3}
\BIBentrySTDinterwordspacing

\bibitem{10.1007/978-3-662-48051-9_2}
M.~M{\"o}ser and R.~B{\"o}hme, ``{Trends, Tips, Tolls: A Longitudinal Study of
  Bitcoin Transaction Fees},'' in \emph{Financial Cryptography and Data
  Security}, M.~Brenner, N.~Christin, B.~Johnson, and K.~Rohloff, Eds.\hskip
  1em plus 0.5em minus 0.4em\relax Berlin, Heidelberg: Springer Berlin
  Heidelberg, 2015, pp. 19--33.

\bibitem{10.1145/2976749.2978408}
\BIBentryALTinterwordspacing
M.~Carlsten, H.~Kalodner, S.~M. Weinberg, and A.~Narayanan, ``{On the
  Instability of Bitcoin Without the Block Reward},'' in \emph{Proceedings of
  the 2016 ACM SIGSAC Conference on Computer and Communications Security}, ser.
  CCS '16.\hskip 1em plus 0.5em minus 0.4em\relax New York, NY, USA:
  Association for Computing Machinery, 2016, p. 154–167. [Online]. Available:
  \url{https://doi.org/10.1145/2976749.2978408}
\BIBentrySTDinterwordspacing

\bibitem{10.1007/978-3-662-45472-5_28}
I.~Eyal and E.~G. Sirer, ``{Majority Is Not Enough: Bitcoin Mining Is
  Vulnerable},'' in \emph{Financial Cryptography and Data Security},
  N.~Christin and R.~Safavi-Naini, Eds.\hskip 1em plus 0.5em minus 0.4em\relax
  Berlin, Heidelberg: Springer Berlin Heidelberg, 2014, pp. 436--454.

\bibitem{DBLP:journals/corr/abs-1112-4980}
\BIBentryALTinterwordspacing
M.~Rosenfeld, ``{Analysis of Bitcoin Pooled Mining Reward Systems},''
  \emph{CoRR}, vol. abs/1112.4980, 2011. [Online]. Available:
  \url{http://arxiv.org/abs/1112.4980}
\BIBentrySTDinterwordspacing

\bibitem{LI2019113094}
\BIBentryALTinterwordspacing
J.~Li, Y.~Yuan, and F.-Y. Wang, ``{A Novel GSP Auction Mechanism for Ranking
  Bitcoin Transactions in Blockchain Mining},'' \emph{Decision Support
  Systems}, vol. 124, p. 113094, 2019. [Online]. Available:
  \url{https://www.sciencedirect.com/science/article/pii/S016792361930123X}
\BIBentrySTDinterwordspacing

\bibitem{houy2014economics}
N.~Houy, ``{The Economics of Bitcoin Transaction Fees},'' \emph{GATE WP}, vol.
  1407, 2014.

\bibitem{8946169}
S.~{Jiang} and J.~{Wu}, ``{Bitcoin Mining with Transaction Fees: A Game on the
  Block Size},'' in \emph{2019 IEEE International Conference on Blockchain
  (Blockchain)}, 2019, pp. 107--115.

\bibitem{6233691}
R.~C. {Merkle}, ``{Protocols for Public Key Cryptosystems},'' in \emph{1980
  IEEE Symposium on Security and Privacy}, 1980, pp. 122--122.

\bibitem{GLOSTEN198571}
\BIBentryALTinterwordspacing
L.~R. Glosten and P.~R. Milgrom, ``{Bid, Ask and Transaction Prices in A
  Specialist Market With Heterogeneously Informed Traders},'' \emph{Journal of
  Financial Economics}, vol.~14, no.~1, pp. 71--100, 1985. [Online]. Available:
  \url{https://www.sciencedirect.com/science/article/pii/0304405X85900443}
\BIBentrySTDinterwordspacing

\bibitem{doi:10.2469/faj.v46.n3.23}
\BIBentryALTinterwordspacing
G.~W. Schwert, ``{Stock Market Volatility},'' \emph{Financial Analysts
  Journal}, vol.~46, no.~3, pp. 23--34, 1990. [Online]. Available:
  \url{https://doi.org/10.2469/faj.v46.n3.23}
\BIBentrySTDinterwordspacing

\bibitem{8751431}
R.~{Banno} and K.~{Shudo}, ``{Simulating A Blockchain Network with SimBlock},''
  in \emph{2019 IEEE International Conference on Blockchain and Cryptocurrency
  (ICBC)}, 2019, pp. 3--4.

\bibitem{litecoin}
\BIBentryALTinterwordspacing
Coinsutra.com, ``{Litecoin},''
  \url{https://coinsutra.com/litecoin-cryptocurrency/}, Last accessed: 13 March
  2020. [Online]. Available:
  \url{https://coinsutra.com/litecoin-cryptocurrency/}
\BIBentrySTDinterwordspacing

\bibitem{vasin2014blackcoin}
P.~Vasin, ``{Blackcoin's Proof-of-Stake Protocol v2},'' \emph{URL:
  https://blackcoin. co/blackcoin-pos-protocol-v2-whitepaper. pdf}, vol.~71,
  2014.

\bibitem{10.1007/978-3-030-51280-4_28}
K.~Karantias, A.~Kiayias, and D.~Zindros, ``{Proof-of-Burn},'' in
  \emph{Financial Cryptography and Data Security}, J.~Bonneau and N.~Heninger,
  Eds.\hskip 1em plus 0.5em minus 0.4em\relax Cham: Springer International
  Publishing, 2020, pp. 523--540.

\bibitem{BitcoinHistoricalData}
\BIBentryALTinterwordspacing
{Kaggle.com}, ``{Bitcoin Historical Data},''
  \url{https://www.kaggle.com/mczielinski/bitcoin-historical-data}, Last
  accessed: 20 Jan 2021. [Online]. Available:
  \url{https://www.kaggle.com/mczielinski/bitcoin-historical-data}
\BIBentrySTDinterwordspacing

\bibitem{TransactionsPerBlock}
\BIBentryALTinterwordspacing
{Blockchain.com}, ``{Transactions per Block},''
  \url{https://www.blockchain.com/charts/n-transactions-per-block}, Last
  accessed: 20 Jan 2021. [Online]. Available:
  \url{https://www.blockchain.com/charts/n-transactions-per-block}
\BIBentrySTDinterwordspacing

\bibitem{Mempool}
\BIBentryALTinterwordspacing
{Blockchain,com}, ``{Mempool},''
  \url{https://www.blockchain.com/charts/mempool-count}, Last accessed: 20 Jan
  2021. [Online]. Available:
  \url{https://www.blockchain.com/charts/mempool-count}
\BIBentrySTDinterwordspacing

\bibitem{TransactionsRatePerSecond}
\BIBentryALTinterwordspacing
{Blockchain.com}, ``{Transactions Rate Per Second},''
  \url{https://www.blockchain.com/charts/transactions-per-second}, Last
  accessed: 20 Jan 2021. [Online]. Available:
  \url{https://www.blockchain.com/charts/transactions-per-second}
\BIBentrySTDinterwordspacing

\end{thebibliography}

\end{document}